      [1999/12/01 v1.4c Il Nuovo Cimento]
\documentclass{cimento}

\usepackage{graphicx,cite}
\usepackage{epsfig}
\usepackage{multirow}

\newcommand{\be}{\begin{equation}}
\newcommand{\ee}{\end{equation}}
\newcommand{\bea}{\begin{eqnarray}}
\newcommand{\eea}{\end{eqnarray}}
\newcommand{\bi}{\begin{itemize}}
\newcommand{\ei}{\end{itemize}}
\newcommand{\ben}{\begin{enumerate}}
\newcommand{\een}{\end{enumerate}}

\newcommand{\lp}{\left(}
\newcommand{\rp}{\right)}

\def\frac#1#2{{{#1}\over {#2}}}
\def\gsim{\mathrel{\rlap{\lower4pt\hbox{\hskip1pt$\sim$}}
    \raise1pt\hbox{$>$}}}         
\def\lsim{\mathrel{\rlap{\lower4pt\hbox{\hskip1pt$\sim$}}
    \raise1pt\hbox{$<$}}}         

\newcommand{\draft}[1]{}

\def\beq{\begin{equation}}  
\def\eeq{\end{equation}}  

\def \n0{N_j^{(0)}}

\def\lapprox{\lower .7ex\hbox{$\;\stackrel{\textstyle <}{\sim}\;$}}
\def\gapprox{\lower .7ex\hbox{$\;\stackrel{\textstyle >}{\sim}\;$}}

\title{Parton Distributions and LHC data}
\author{J.~Rojo\from{ins:x}}
\instlist{\inst{ins:x} Dipartimento di Fisica, Universit\`a di Milano and INFN, Sezione di Milano, Via Celoria 16, I-20133 Milano, Italy}

\PACSes{12.38.-t,12.38.Lg}

\begin{document}

\maketitle

\begin{abstract}
In this contribution we briefly report on the progress
and open problems in parton distribution functions (PDFs),
with emphasis on their implications for LHC phenomenology.
Then we study the impact of the recent ATLAS and CMS $W$ lepton asymmetry data
on the NNPDF2.1 parton distributions. We show that these data
provide the first constrains on PDFs from LHC measurements.
\end{abstract}

\section{Progress and open problems in parton distributions}

The quantitative control of
the Standard Model contribution to
 collider signal and background processes at the
few percent level is a necessary ingredient not
only for precision physics, but also for discovery at the
LHC. The precision determination of parton distribution functions (PDFs) 
is essential in order to achieve this level of theoretical accuracy.

There has been substantial progress in PDF analysis
in the last years, and it is thus impossible to review it
in detail in this contribution.
A recent concise report of the status of the field
 can be found in Ref.~\cite{higgsYR}, while  more
detailed reviews can be found in Refs.~\cite{Forte:2010dt,Campbell:2006wx,DeRoeck:2011na,Alekhin:2011sk}. 
In this contribution we restrict ourselves to highlight some
important topics in PDF determinations. First of all,  we will 
sketch the current status
of PDF fits and discuss some of the open problems in the field.
Then we will discuss how the ATLAS and
CMS measurements of the $W$ lepton asymmetry data provide
the first constraints on PDFs from the LHC, thus paving the way for PDFs
based on LHC data.

PDF analysis have entered the era in which they 
can be considered as a quantitative
science. 
An ideal PDF determination should satisfy several
important requirements~\cite{Forte:2010dt}. These include being based on a dataset which is as wide as possible, in
  order to ensure that all relevant experimental information is
  retained,  to use  a sufficiently general and unbiased parton
  parametrization and to provide statistically consistent confidence
levels for PDF
uncertainties. Moreover, such ideal set should include 
heavy quark mass effects through a GM-VFN
  scheme~\cite{lhhq} and be based on computations 
performed at the highest available perturbative
  order.
Finally, PDF sets should be provided
 for a variety of values of $\alpha_s$,
  reasonably thinly spaced,  similarly
 for the heavy quark masses, and should include an estimate of 
uncertainties related to the
  truncation of the perturbative expansion. While for each of these
aspects there has been sizable progress in the recent years, still
no PDF sets fulfills all these conditions.

One important development in PDFs in the recent years has been
the NNPDF approach~\cite{DelDebbio:2004qj,DelDebbio:2007ee,Ball:2008by,Ball:2009mk,Ball:2010de}.
Thanks to a combination of Monte Carlo techniques and the use of artificial neural networks, the NNPDF approach avoids some of the drawbacks of
the standard approach like the bias due to the arbitrary choice of input functional forms or the use of
linear approximations for PDF uncertainty estimation.
The most updated NNPDF set is NNPDF2.1~\cite{Ball:2011mu}, an unbiased NLO 
global
fit of all relevant hard scattering data based on the FONLL-A
GM-VFN scheme~\cite{Forte:2010ta}.

Several groups provide regular updates of their PDF sets: in alphabetic
order these are ABKM, CT, HERAPDF, JR, MSTW and NNPDF.
In Table~\ref{tab:PDFsummary} we summarize some of the features of
the most updated PDF sets from each collaboration. We consider only 
those sets available in the LHAPDF library. We compare the dataset,
parametrization, method to estimate PDF uncertainties,
perturbative order at which PDFs are available, the theoretical schemes
adopted to include heavy quark mass effects and the treatment of
 the strong coupling $\alpha_s$. More details on each of these
issues can be found in Refs.~\cite{Forte:2010dt,DeRoeck:2011na,Alekhin:2011sk},
as well as in the original publications of each group.

\begin{table}[t]
  \caption{Summary of the features of the most updated PDF sets from
each group.
The CT10, MSTW08 and NNPDF2.1 sets include data from a wide variety of
physical processes and are thus called {\it global} PDF sets. See text
for more details. \label{tab:PDFsummary}}
\begin{tabular}{|c||c|c|c|c|}
\hline
  & Ref & Dataset & Parametrization & PDF uncertainties  \\
\hline
\hline
ABKM09 & \cite{Alekhin:2009ni}   & DIS+DY & Polynomial  &  Hessian, standard tol. \\
CT10    & \cite{Lai:2010vv} & DIS+DY+W/Z+jet & Polynomial & Hessian, dyn. tol.   \\
HERAPDF1.0 & \cite{heracombined:2009wt} & DIS & Polynomial  &  Hessian, standard tol.\\
JR08     & \cite{JimenezDelgado:2008hf} & DIS+DY+jet& Polynomial  &  Hessian, fixed tol. \\
MSTW08   & \cite{Martin:2009iq} & DIS+DY+W/Z+jet & Polynomial  &  Hessian, dyn. tol.  \\
NNPDF2.1 & \cite{Ball:2011mu} & DIS+DY+W/Z+jet & Neural Nets  & Monte Carlo   \\
\hline
\end{tabular}

\begin{tabular}{|c||c|c|c|}
\hline
   & PT order & Heavy Quarks  & Strong coupling \\
\hline
\hline
ABKM09    & NLO/NNLO & FFNS  & Fitted   \\
CT10      & NLO & S-ACOT-$\chi$  & Fixed + range of values   \\
HERAPDF1.0  & NLO & TR &  Fixed  \\
JR08      & NLO/NNLO &  FFNS &  Fitted  \\
MSTW08    & LO/NLO/NNLO & TR & Fitted + range of values   \\
NNPDF2.1  & NLO & FONLL-A  & Fixed + range of values  \\
\hline
\end{tabular}
\end{table}

The main difference arises from the data sets used in each of the various
analysis. The CT10, MSTW08 and NNPDF2.1 sets include data from a wide variety of
physical processes and are thus called {\it global} PDF sets. Other PDF sets
use more restrictive subsets, like ABKM09, which excludes Tevatron jet and
weak vector production data and HERAPDF1.0, that is based solely on HERA data.

PDFs are typically parametrized with relatively simple functional
forms like $q(x,Q_0^2)\sim x^{a}(1-x)^{b}P(x,c,d,\ldots)$ with $P$ a polynomial
that interpolates between the small and large--$x$ regions. These unjustified
theoretical 
assumptions introduce a potentially large functional form bias in PDF
determinations. 
The NNPDF approach bypasses this problem using
neural networks as universal unbiased interpolants. Related techniques
for general PDF parametrizations like Chebishev polynomials have also
been discussed in the literature~\cite{Glazov:2010bw,Pumplin:2009bb}.

PDF uncertainties are estimated by all groups (but NNPDF) using the Hessian
method. However,  different choices for the tolerance $T=\sqrt{\Delta\chi^2}$
adopted to define 1--sigma PDF uncertainties are used. 
For example, while HERAPDF1.0
and ABKM08 are based on a textbook tolerance $\Delta\chi^2=1$, MSTW08
and CT10 adopt a dynamical tolerance criterion that results in
tolerances $\Delta\chi^2\ge1$, which are moreover different for each
eigenvector direction. The need for large tolerances has been
suggested to partly arise when 
restrictive input functional forms are used~\cite{Pumplin:2009bb}.
NNPDF, on the other hand, is based on 
the Monte Carlo approach, that is, a sampling in the
space of experimental data, that allows an exact uncertainty propagation
from data to PDFs and from these to physical observables.

\begin{figure}[t]
\caption{\small Comparison of the NLO total cross sections
for $W^+$ and $t\bar{t}$ production and their
combined PDF+$\alpha_s$ uncertainties at the LHC 7 TeV between
the most updated PDF sets of each group. Plots from G. Watt.
 \label{fig:comp7tev}} 
\begin{center}
\epsfig{width=0.49\textwidth,figure=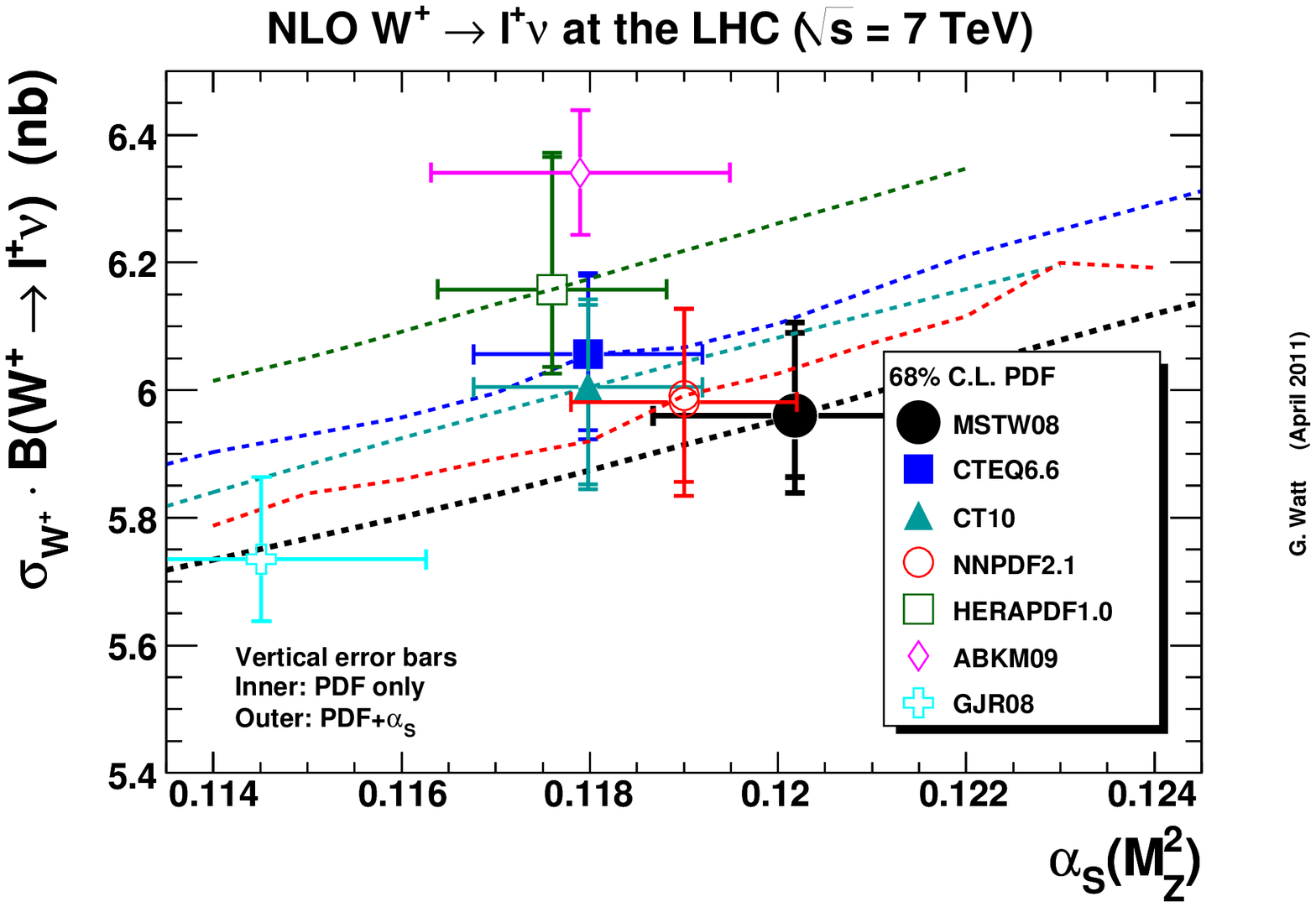}
\epsfig{width=0.49\textwidth,figure=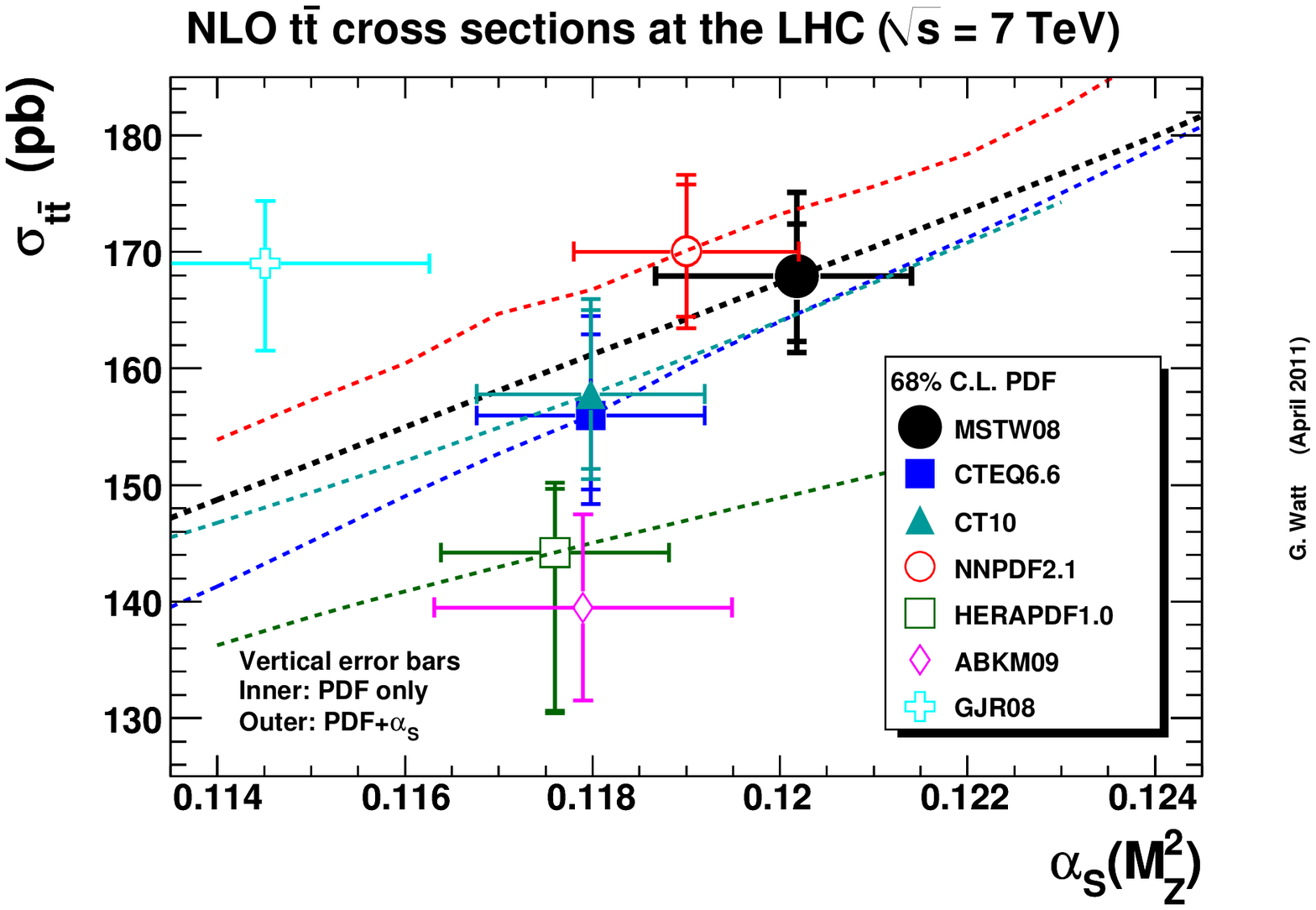}
\end{center}
\end{figure}

Recently, a detailed benchmarking of the predictions for
relevant LHC observables from modern NLO PDF sets was performed in the
context of the PDF4LHC working group~\cite{Alekhin:2011sk}. 
 In Fig.~\ref{fig:comp7tev} we compare the NLO predictions for different PDF sets
for two important LHC observables, the
 total $W^+$ and $t\bar{t}$ 
cross sections.  One of the conclusions from that study is that the agreement 
between {\it global} PDF sets is reasonable for most
LHC processes, much better than for sets based on 
restrictive datasets. However, it was also clear that even
within global sets there are important discrepancies whose origin
needs still to be understood, related for example to the large-$x$ gluon and
to strangeness. Another recent benchmark study, this time at NNLO, was
presented in~\cite{Alekhin:2010dd}.

\begin{figure}[t]
\caption{\small Left plot: Comparison of the NLO Higgs production
cross section with the combined PDF+$\alpha_s$ uncertainties from NNPDF2.0,
MSTW08 and CTEQ6.6, and the resulting PDF4LHC recipe~\cite{Botje:2011sn} envelope, from 
Ref.~\cite{higgsYR}. Right plot: comparison of the MSTW08 and
the preliminary NNPDF2.1 NNLO predictions for the NNLO Higgs production cross section. For the MSTW08 prediction two values of $\alpha_s$ have been used.
 \label{fig:comphiggs}} 
\begin{center}
\epsfig{width=0.43\textwidth,figure=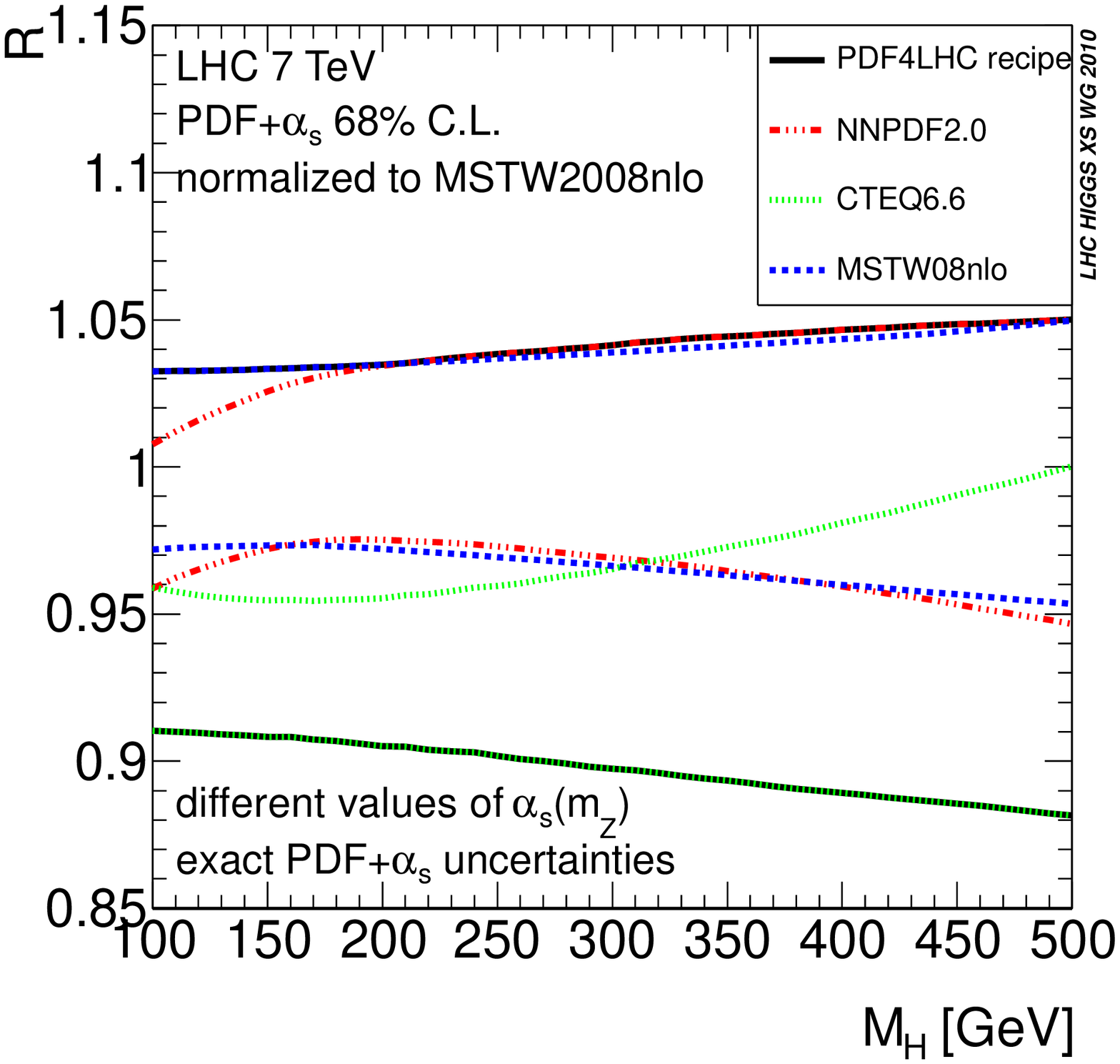}
\epsfig{width=0.56\textwidth,figure=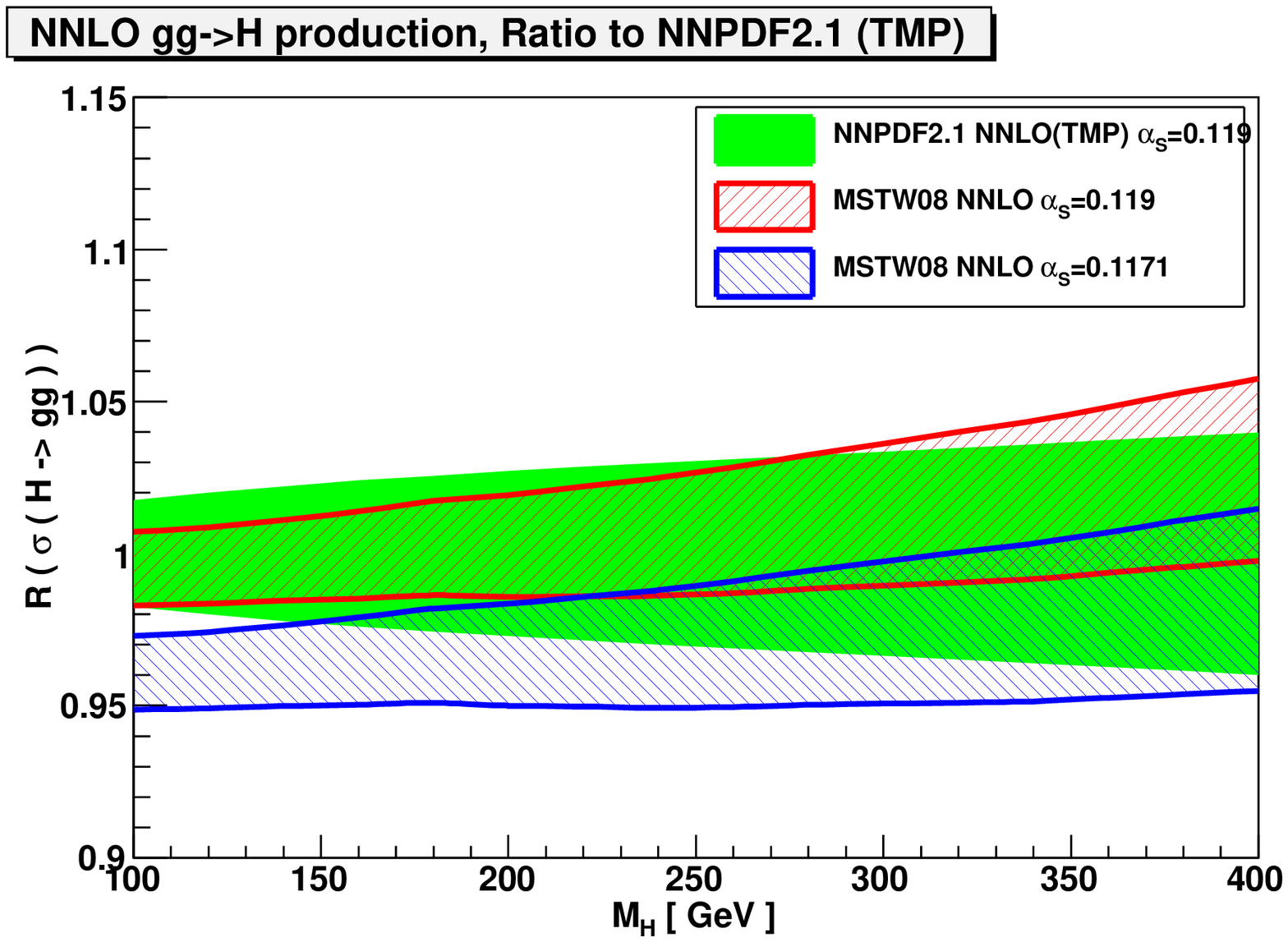}
\end{center}
\end{figure}

The PDF4LHC exercise allowed
to elucidate differences and similarities between PDF sets. In particular
it showed that the most important source of difference between sets is 
the choice
of fitted data. 
This study was the basis of the PDF4LHC recommendation~\cite{Botje:2011sn},
that suggests to take the envelope of the combined PDF+$\alpha_s$ uncertainties
from the three global PDF sets, CT10, MSTW08 and NNPDF2.1, to estimate
the PDF+$\alpha_s$ uncertainty on LHC processes. The PDF4LHC has
been adopted by ATLAS and CMS in those analysis sensitive to
PDFs, and in particular 
the LHC Higgs cross section
working group~\cite{higgsYR} uses the PDF4LHC recipe to estimate the combined
PDF+$\alpha_s$ uncertainty in their theoretical predictions, see
Fig.~\ref{fig:comphiggs}. 
The same recipe has been used to derive 
the most updated Tevatron Higgs exclusion limits~\cite{Aaltonen:2011gs}.

Let us now turn to discuss some open problems in PDF fits: the treatment
of  $\alpha_s$, Higgs production at hadron colliders and deviations from
DGLAP in HERA data.
The treatment of the strong coupling in PDF fits is a source of differences
between sets, as summarized in Table~\ref{tab:PDFsummary}. 
Some groups, like MSTW or ABKM, determine $\alpha_s$
simultaneously with the PDFs, while others, like CT or NNPDF,
take for $\alpha_s$ a fixed value close to the
PDG average~\cite{Nakamura:2010zzi},
$\alpha_s\lp M_Z\rp=0.1184\pm 0.0007$ in the
latest update. Differences between PDF sets are reduced when
a common value of $\alpha_s$ is used, as shown also in the comparison
plots of Fig.~\ref{fig:comp7tev}.

Let us emphasize that the choice of fixing $\alpha_s$ to the PDG value
in the reference PDF set 
is not necessarily related to the sensitivity of a given
PDF analysis 
to $\alpha_s$. Rather, it reflects the idea than the average of 
$\alpha_s$ from a wide
range of processes, including some like $\tau$ decays unrelated to the proton
structure, is necessarily more accurate than the determination from a single
PDF fit.  For example, NNPDF~\cite{Lionetti:2011pw}
has recently performed a NLO determinations of the strong coupling, finding good consistency
with the PDG value: $\alpha_s\lp M_Z\rp=0.1191 \pm 0.0006$, 
where the uncertainty is purely statistical.

The treatment of $\alpha_s$ is closely related to 
one of the most important
process at the LHC, the Higgs production cross section in its 
dominant production channel
of gluon fusion.
This process
is very sensitive to $\alpha_s$~\cite{Demartin:2010er}, since
the partonic cross section depends as $\mathcal{O}\lp \alpha_s^2\rp$ 
already at leading order, and has received a lot of attention
 recently
due to claims that theoretical uncertainties were being underestimated.
Preliminary NNLO results from NNPDF, shown in Fig.~\ref{fig:comphiggs}, 
 suggest a reasonable agreement with the MSTW08 NNLO prediction, 
as was already the case
at NLO, thus confirming the PDF4LHC recipe estimates. 
It is also clear how the use of a common
value of $\alpha_s$ improves further the agreement between the two 
sets.

Another open problem in PDF determinations are the potential
departures from fixed-order DGLAP evolution in small-$x$ and
$Q^2$ HERA data. The analysis of Refs.~\cite{Caola:2009iy,Caola:2010cy} 
found evidence for deviations
from NLO DGLAP in the small-$x$ combined HERA-I data, consistent
with small-$x$ resummation and non-linear dynamics but not with NNLO
corrections.
This effect has been confirmed by the HERAPDF analysis, which also
finds a worse fit quality at NNLO for the small-$x$ data. 
A related CT10~\cite{Lai:2010nw} analysis found some hints as well but it was restricted to the
use of few functional forms for the small-$x$ PDFs. If deviations
from DGLAP for low-$x$ HERA data are confirmed, this 
suggests that small-$x$ resummation~\cite{Forte:2009wh} is a necessary 
ingredient 
in order to use
all the potential of  HERA data for precision LHC physics.

\section{Constraining PDFs with LHC $W$ asymmetry data}

\begin{figure}[t]
\caption{\small The kinematic coverage in the $(x,Q^2)$ plane
for $W$ production at the LHC in the central (ATLAS and CMS)
and forward (LHCb) regions. \label{fig:wlasy}}
\begin{center}
\epsfig{width=0.49\textwidth,figure=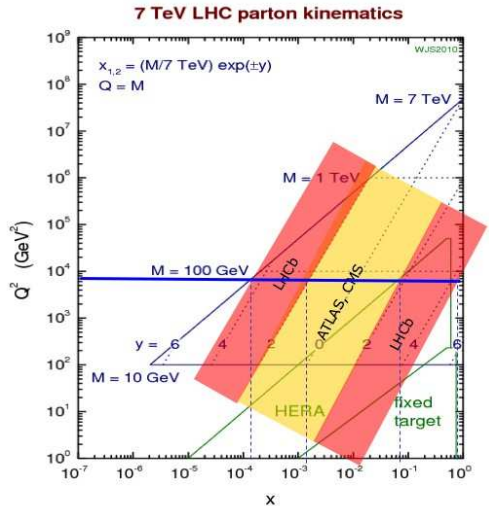}
\end{center}
\end{figure}

We now turn to discuss the first constraints on PDFs from
LHC data, provided by the ATLAS~\cite{Aad:2011yn} and CMS~\cite{Chatrchyan:2011jz} measurements of
the leptonic $W$ asymmetry\footnote{There exist as well
preliminary data from LHCb that will be sensitive to even smaller
and larger values of $x$, see Fig.~\ref{fig:wlasy}.}. 
As it is well known,
$W$ production at hadron colliders is sensitive to the light quark
and antiquark PDFs at medium and small-$x$. The kinematic
coverage of $W$ production at the LHC is summarized in 
Fig.~\ref{fig:wlasy}. We have studied the impact of
the $W$ asymmetry data using the Bayesian reweighting method of Ref.~\cite{Ball:2010gb}.  
Bayesian reweighting is a powerful technique to efficiently determine
the impact of new data into PDFs without the need of refitting.
This method also allows to determine
the internal consistency of the data sets and their compatibility 
with the global fit.

A detailed discussion of the impact of LHC data
on NNPDF will be presented elsewhere. 
In this contribution we restrict ourselves to some selected preliminary results.
We will show results for the impact of the
combined ATLAS and CMS data. In the case of CMS we consider
the more inclusive dataset (with the cut in lepton transverse
momentum of $p_t^l\ge$ 25 GeV) and both electrons
and muons.  For ATLAS only the muon asymmetry has been presented.
The theoretical predictions have been computed with the
DYNNLO generator~\cite{Catani:2009sm} at NLO accuracy for 
NNPDF2.1. The kinematic cuts are the same as in
the respective experimental analyses.

\begin{figure}[t]
\caption{\small The ATLAS and CMS $W$ lepton asymmetry data compared to the NNPDF2.1 predictions before
and after reweighting. \label{fig:wasy-rw}}
\begin{center}
\epsfig{width=0.49\textwidth,figure=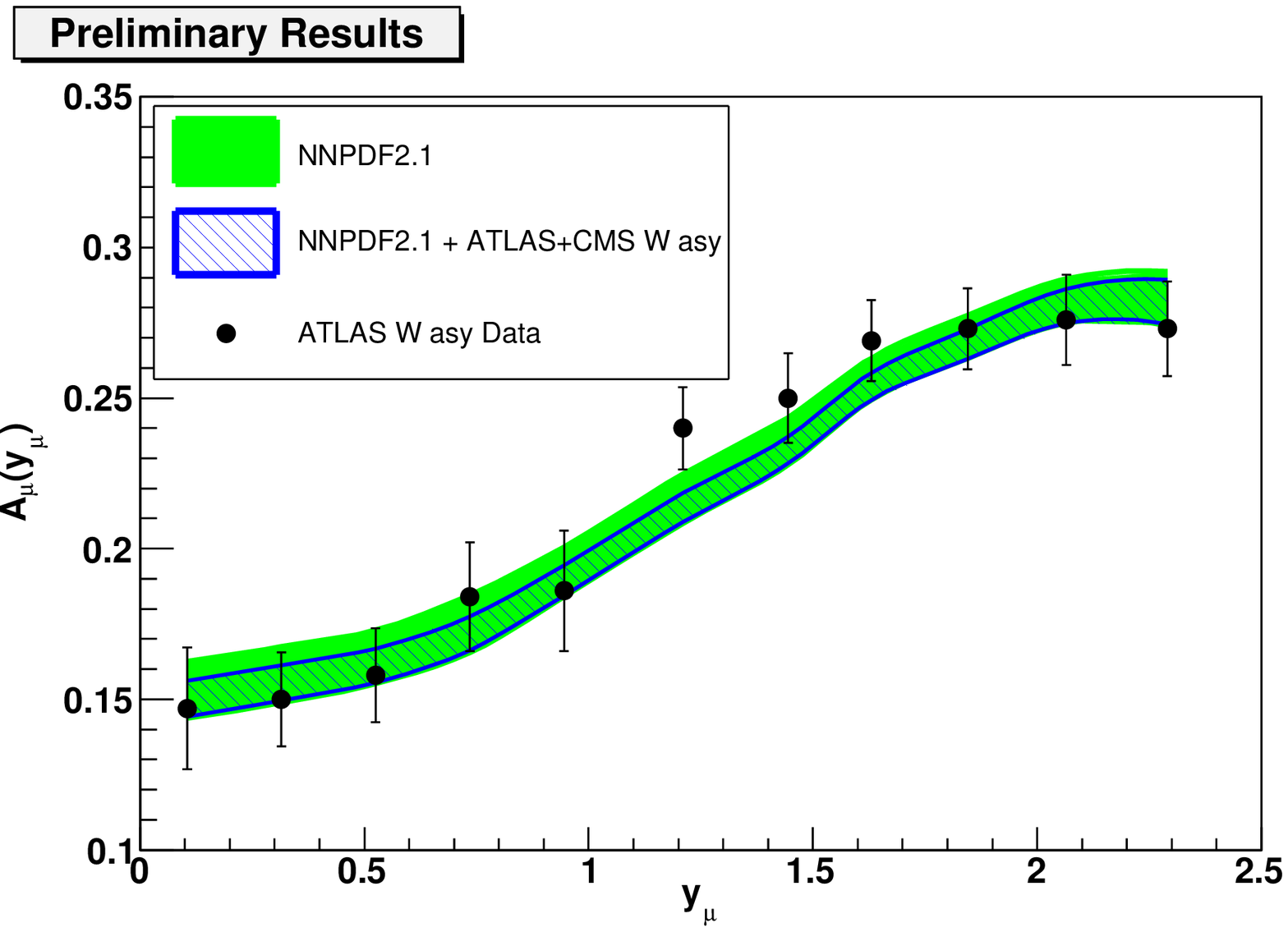}
\epsfig{width=0.49\textwidth,figure=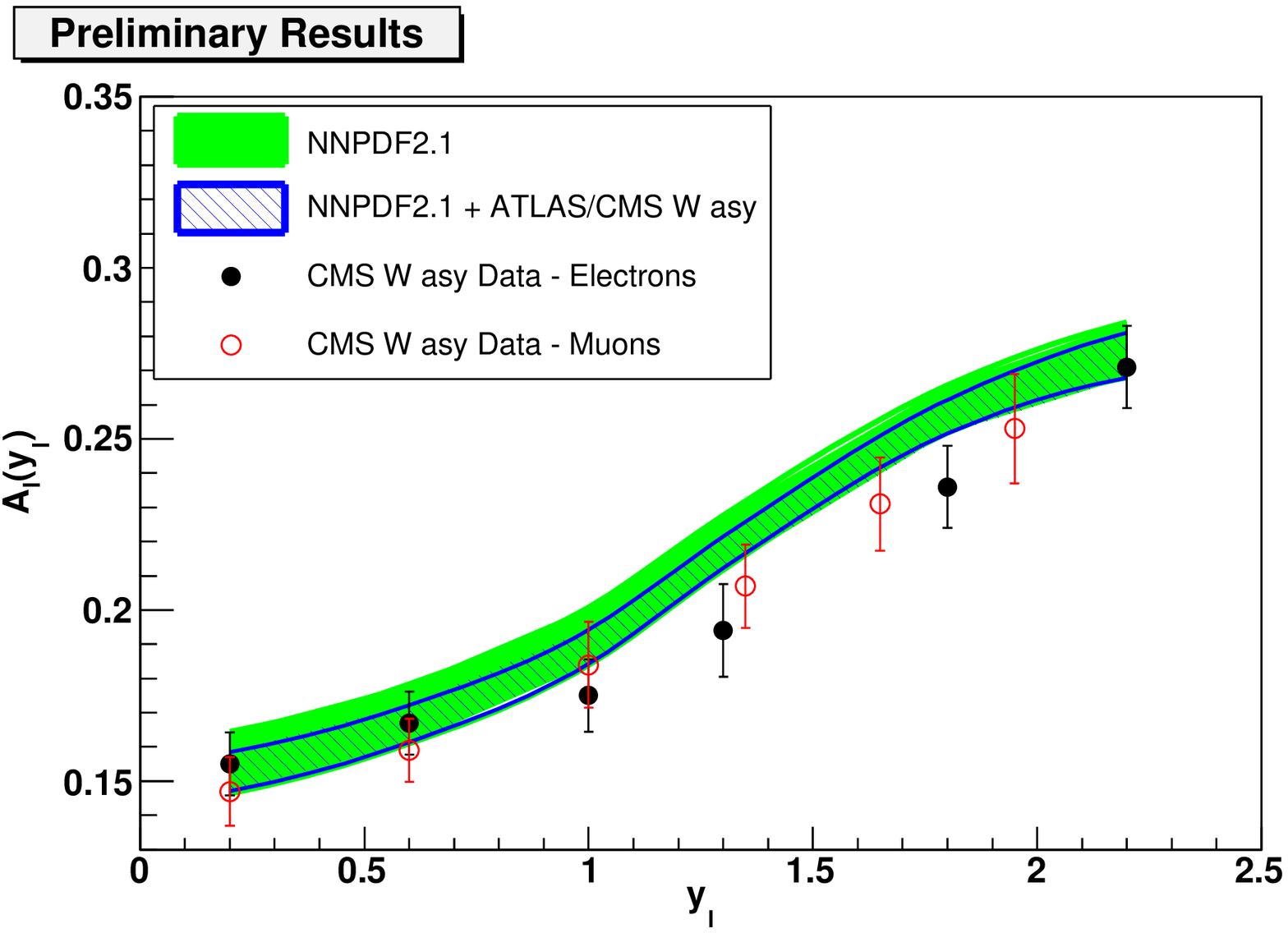}
\end{center}
\end{figure}

In Fig.~\ref{fig:wasy-rw} we compare the ATLAS and CMS lepton
asymmetry data with the NNPDF2.1 predictions before and after
including the effect of these data sets. We notice
that the data is already nicely consistent with the NNPDF2.1 prediction
within the respective uncertainties. 
After including the LHC measurements, one finds that the
$W$ asymmetry data constraints
the PDF uncertainties and leads to an even better agreement with the
data. A more detailed statistical analysis confirms that
the ATLAS and CMS data are consistent between them and with the
experiments included in the global PDF analysis. After reweighting, the
$\chi^2$ per data point of the combined CMS and ATLAS data is $\sim 1$.

\begin{figure}[t]
\caption{\small The impact of the ATLAS and CMS lepton asymmetry data on the
relative uncertainty of
the light quark and antiquark NNPDF2.1 PDFs.
\label{fig:wasy-pdfs}}
\begin{center}
\epsfig{width=0.49\textwidth,figure=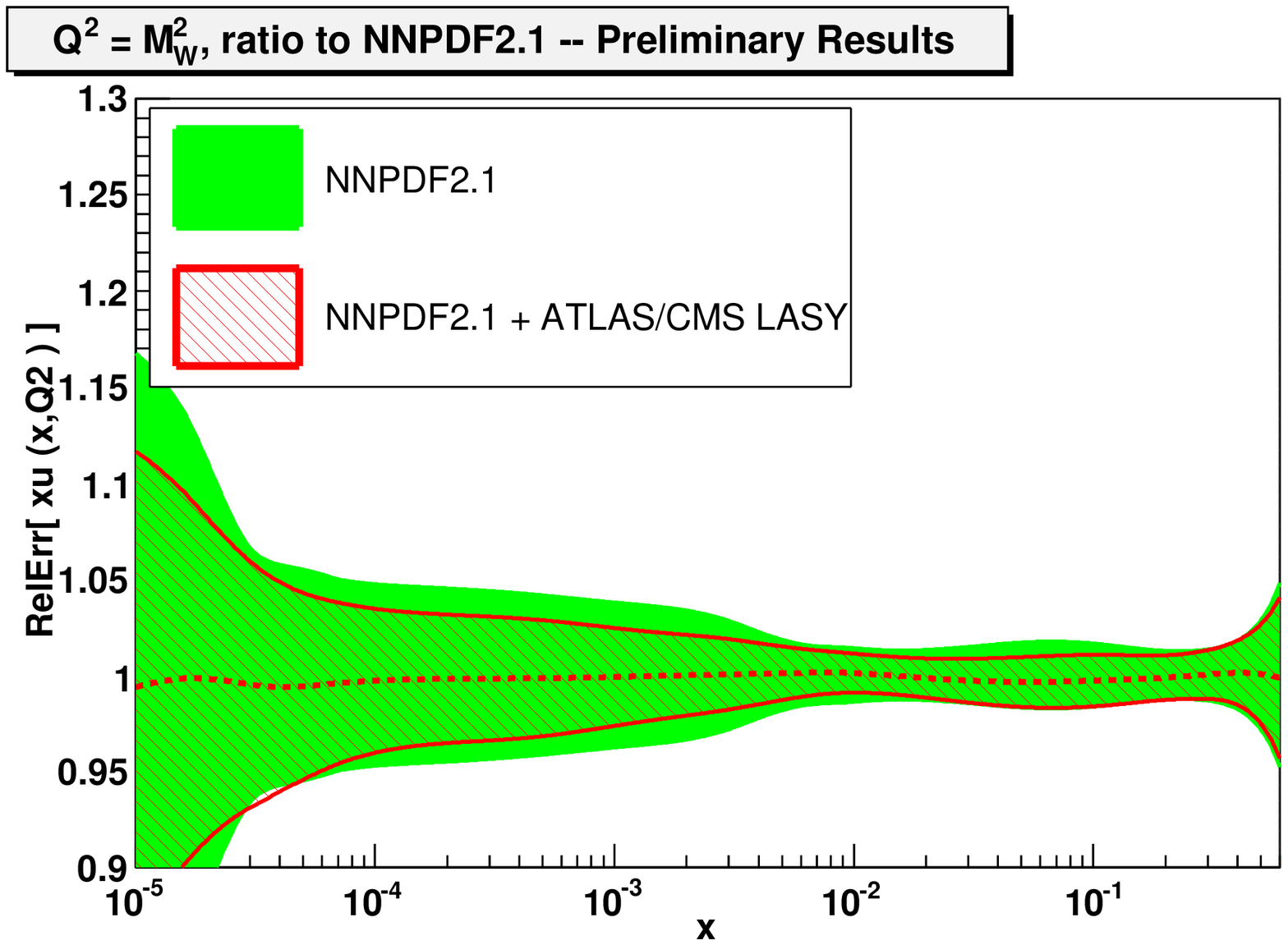}
\epsfig{width=0.49\textwidth,figure=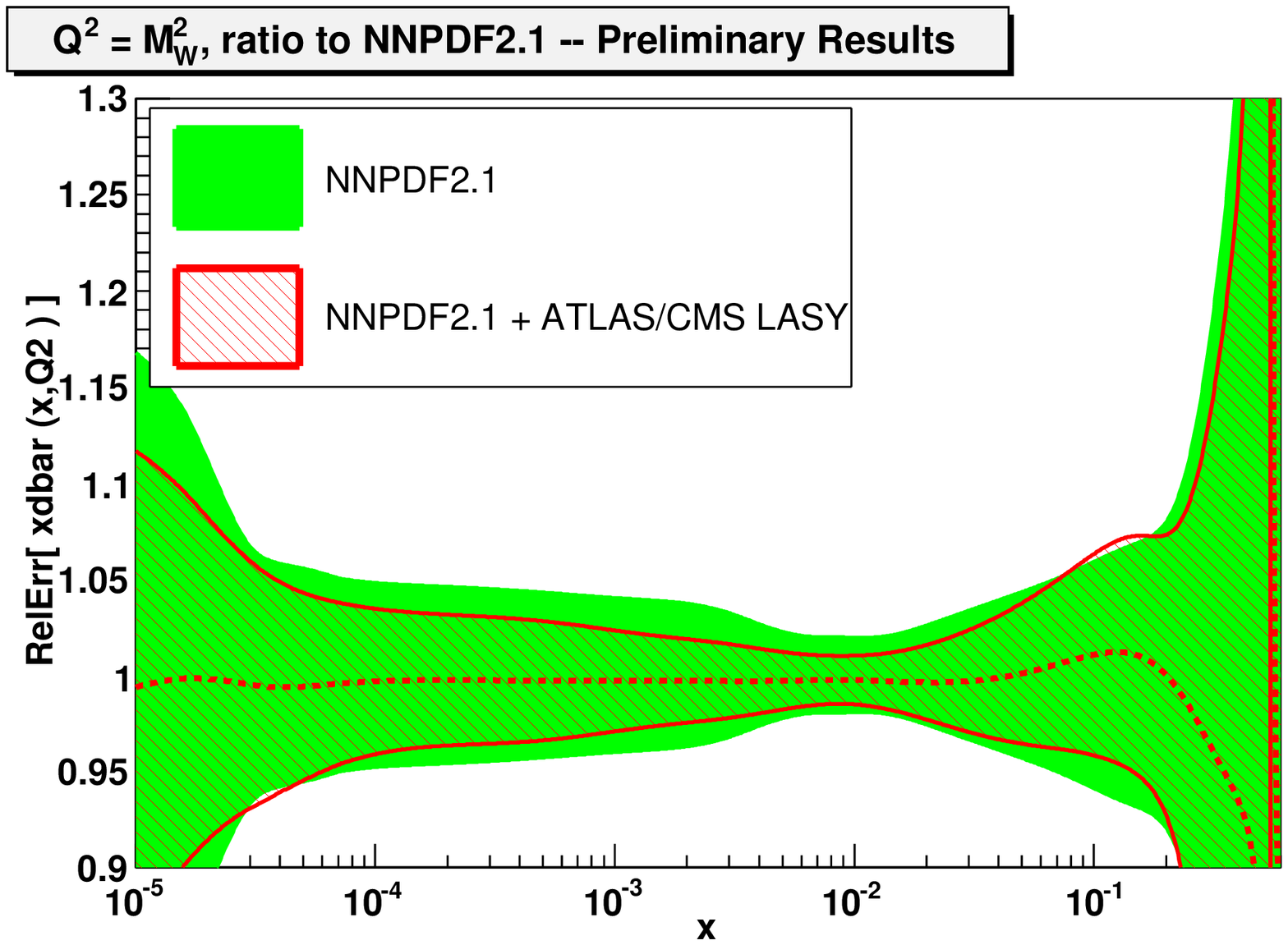}
\end{center}
\end{figure}

Next, in Fig.~\ref{fig:wasy-pdfs} we show the constraints on the PDFs provided by the
combined ATLAS and CMS $W$ asymmetry data.  
We find that the PDF uncertainties are reduced for
medium and small-$x$ light quark and antiquarks, by a factor 
that can be as large
as $\sim 30$--$40$\%. The impact on other PDFs is smaller. The central
PDF prediction is almost unaffected by the LHC data, confirming further
the consistency of the $W$ asymmetry measurements with the global fit.
At large-$x$ the constrains
are weaker, as expected from the kinematic coverage shown in 
Fig.~\ref{fig:wlasy}.
Upcoming measurements of this asymmetry by LHCb might help in reducing PDF
uncertainties in the large-$x$ region.

Note that these preliminary results have been derived from a 
sample of only $N_{\rm rep}=100$
Monte Carlo replicas.  This means that there can be non--negligible fluctuations and explains why 
PDF uncertainties are apparently reduced
even at very small-$x$, outside the kinematic coverage
 of the ATLAS and CMS data.

To summarize, we have shown that
the $W$ lepton asymmetry is the first dataset from the LHC that has the
 precision to constrain
PDFs and thus improve the accuracy of Standard Model computations for
LHC processes. 
We have quantified this impact on the light quark and antiquark
PDFs, and found that PDF uncertainties can be reduced by factors
up to $\sim 40\%$ at medium and small-$x$.
More constrains on PDFs should soon be available  from 
upcoming LHC measurements.

\section{Outlook}

In this contribution we have briefly reviewed recent developments
and open problems related to PDFs, with emphasis on
their implications for the LHC physics program. While our understanding
of the proton structure has seen a huge progress in the recent years, there
are still open questions that need to be answered, and that are important
to improve even further the accuracy of theoretical predictions at
the LHC.  We have also
presented preliminary results on the impact of the LHC $W$ 
lepton asymmetry data on the NNPDF2.1 set. We have shown that these data provide the first
constraints on PDFs from LHC measurements, in particular they help to pin down
with better accuracy the medium and small $x$ light quarks and 
antiquarks.

In the medium term, LHC measurements will provide very important
constraints on most  PDF combinations. This will allow
 parton distributions to be derived solely
from collider data: HERA, Tevatron and the LHC. Collider data
is more robust theoretically and experimentally than
low-energy fixed target data, that now provide basic
constrains in global PDF analysis. In order to achieve
this program, several measurements will be provided by the LHC:
 $Z$--boson rapidity distributions, low mass Drell-Yan differential
distributions, high $E_T$ jets and photons, and $W/Z$ production in
association with heavy quarks. The increased experimental 
and theoretical accuracy on PDFs determined this way will provide a solid
ground for precision Standard Model predictions and searches for 
new physics at the LHC.

\acknowledgments
I would like to thank the La Thuile 2011 organizers for their kind invitation
to present this review. I thank all the members of the NNPDF Collaboration
for endless discussions on PDFs, specially S. Forte. I thank G. Watt
for providing the plots in Fig.~\ref{fig:comp7tev}.

\end{document}